\documentclass[final,3p,times,twocolumn]{elsarticle}
 \biboptions{comma,sort&compress}
\usepackage{here}
 \usepackage{graphicx}
  \usepackage{epsfig}
\def\nin{\noindent}
\def\beq{\begin{equation}}
\def\eeq{\end{equation}}
\def\bea{\begin{eqnarray}}
\def\eea{\end{eqnarray}}

\journal{Nuc. Phys. (Proc. Suppl.)}

\begin{document}

\begin{frontmatter}



\title{Thermodynamics of AdS/QCD within the 5D dilaton-gravity model}

 \author[label1]{E.~Meg\'{\i}as\corref{cor1}}
  \address[label1]{Institute for Theoretical Physics, University of Heidelberg, Germany}
\cortext[cor1]{Speaker}
\ead{emegias@tphys.uni-heidelberg.de}

 \author[label1,label2]{H.J.~Pirner}
  \address[label2]{Max Planck Institute for Theoretical Physics, Heidelberg, Germany}
%\cortext[label3]{Supported by FAPESP within the France-Brazil program.}
\ead{pir@tphys.uni-heidelberg.de}

\author[label1]{K.~Veschgini}
\ead{K.Veschgini@tphys.uni-heidelberg.de}

\begin{abstract}
\noindent
We calculate the pressure, entropy density, trace anomaly and speed of
sound of the gluon plasma using the dilaton potential of
Ref.~\cite{Galow:2009kw} in the dilaton-gravity theory of AdS/QCD. The
finite temperature observables are calculated from the Black Hole
solutions of the Einstein equations, and using the Bekenstein-Hawking
equality of the entropy with the area of the horizon.  Renormalization
is well defined, because the $T=0$ theory has asymptotic
freedom. Comparison with lattice simulations is made.

\end{abstract}

\begin{keyword}
QCD thermodynamics \sep Gauge-gravity correspondence \sep Black Holes

\end{keyword}

\end{frontmatter}

% \linenumbers

\section{Introduction}
\label{sec:introduction}
\nin
%%%%%%%%%%%%
The duality of string theory with 5-dimensional gravity is nowadays a
powerful tool to study the strong coupling properties of gauges
theories and in particular of QCD, either at zero or finite
temperature. In conformal $AdS_5$ the metric is well known. It has a
horizon in the bulk space at $r_T= \frac{\pi \ell^2}{ \beta}$ where
$\beta=1/T$ and $\ell$ is the size of the AdS-space.  Conformal
solutions for entropy scale like $s \propto T^3$, since the
3-dimensional area of the horizon is given as $A \propto 2 \pi^2
r_T^3$. Promising solutions of this conformal theory have been
proposed to the problem of viscosity $\eta$~\cite{Policastro:2001yc}
with a small constant value for $\eta/s$. Top to bottom approaches
based on the conformal SYM with fermions have been investigating the
chiral phase transition and problems at finite density
\cite{Erdmenger:2007cm,Mateos:2006nu}.

 To extend this duality to SU($N_c$) Yang-Mills theory, one of the first tasks is to control the breaking of conformal invariance. In this work we will studied a 5D dilaton-gravity model introduced in~\cite{Gursoy:2008za}, with the action
\begin{eqnarray}
{\cal S} &=& \frac{1}{16 \pi G_5}\int d^5x \sqrt{-G}\left(R-\frac{4}{3}\partial
_{\mu }\phi \partial ^{\mu }\phi -V(\phi )\right) \nonumber \\
&& -\frac{1}{8 \pi G_5}
\int _{\partial M} d^4x \sqrt{-H} K  \,. \label{eq:action5D}
\end{eqnarray}
The last term is the Gibbons-Hawking term, with $K$ being the extrinsic curvature and $H$ the induced metric on the boundary $\partial M$. The dilaton potential~$V(\phi)$ is related to the $\beta$-function in a one-to-one relation. While the $\beta$-function is well known at high energies, there is no general consensus about its IR behavior. We assume the following parameterization
\begin{eqnarray}
  \beta(\alpha) &=& -b_2\alpha + \bigg[b_2\alpha  +
\left(\frac{b_2}{\bar{\alpha}}-\beta_0\right)\alpha^2 \nonumber \\
&&\quad +
\left(\frac{b_2}{2\bar{\alpha}^2} - \frac{\beta_0}{\bar{\alpha}}-\beta_1 \right)
\alpha^3 \bigg] e^{-\alpha/\bar{\alpha}}\,,
\label{eq:beta}
\end{eqnarray}
which was proposed in Ref.~\cite{Galow:2009kw} for the computation of the heavy $Q\bar{Q}$ potential at zero temperature within this dilaton-gravity model. This parameterization has the standard behavior in the UV region limit $\beta(\alpha) \simeq -\beta_0 \alpha^2 - \beta_1 \alpha^3 + \dots$ for $\alpha \ll \bar{\alpha}$, and generates confinement in the IR region, where it behaves as $\beta(\alpha) \simeq -b_2  \alpha$ for $\alpha \gg \bar{\alpha}$. The optimum values to reproduce the $Q\bar{Q}$ potential at $T=0$ are~\cite{Galow:2009kw} 
\begin{equation}
b_2 = 2.3  \,, \quad \bar\alpha = 0.45 \,, \quad \ell = 4.389 \textrm{GeV}^{-1} \,. \label{eq:b2alpha}
\end{equation}
This parametrization also allows to obtain the running QCD-coupling $\alpha(E)$ in the $\overline{\textrm{MS}}$-scheme from the running of the dilaton $\phi(z)$ in the bulk by mapping the bulk coordinate $z$ to the energy scale in Ref.~\cite{Galow:2009kw}. Small values of the bulk coordinate $z$ correspond to large energies and large values of $z$ describe the infrared physics.
%%%%%%%%%%%%
\section{Equations of motion}
\label{sec:eq_mot}
\nin
%%%%%%%%%%%%%%%%%%%%%
The dilaton-gravity model of Eq.~(\ref{eq:action5D}) has two different types of solutions for the metric. The thermal gas solution corresponds to the confined phase, and the black hole solution characterizes the deconfined phase and it has a horizon localized in the bulk coordinate at $z=z_h$ similar to the situation in 4-dim gravity~\cite{Gursoy:2008za}. We have studied in Ref.~\cite{Veshgini:2009nw} the heavy $Q\bar{Q}$ free energy in the confined phase of QCD by using a thermal gas metric. The black hole metric takes the form
\begin{eqnarray}
ds^2       &=& b^2(z)\left(f(z) d\tau^2 + dx_kdx^k + \frac{dz^2}{f(z)}\right) \,, \label{eq:bhmetric}
\end{eqnarray}
where $z = \ell^2/r$ and $f(z)$ has the properties
\begin{eqnarray}
f(z_h) = 0 \,, \qquad f'(z_h) = -4\pi T \,. \label{eq:fprop}
\end{eqnarray}
The second expression in Eq.~(\ref{eq:fprop}) follows from regularity of the metric at the horizon. This solution only exists at high enough temperatures~\cite{Gursoy:2008za}. The classical equations of motion corresponding to the BH solutoin read~\cite{Gursoy:2008za,Alanen:2009xs,Megias:pr}
\begin{eqnarray}
&&W' = \frac{16}{9} b W^2 - \frac{1}{f} \left( W f' - \frac{3}{4} b V \right) \,, \label{eq:em1} \\
&&b' = -\frac{4}{9} b^2 W \,, \label{eq:em2} \\
&&\alpha' = \alpha \sqrt{b W'} \,, \label{eq:em3} \\ 
&&f'' = \frac{4}{3} f' b W \label{eq:em4} \,,
\end{eqnarray}
where the {\it superpotential} is defined as $W(z) = -9b'(z)/(4 b^2(z))$.

\section{Thermodynamics}
\label{sec:thermodynamics}
%\nin
%%%%%%%%%%%%%%%%%%%%%%%%%%%
The phase transition from the confined glueball gas to the deconfined
gluon plasma can be qualitatively understood as follows. At low
temperatures the infrared physics is determined by the thermal gas
metric with $f =1$. At a minimal temperature the black hole solution
appears with a horizon at $z_h$ and $f \neq 1$ and becomes the
preferred solution at the phase transition which is first order.  The
phase transition temperature is fully fixed by the scales of the
theory at zero temperature. The position of the black hole horizon
intervenes between the confinement region in bulk space and determines
from then on the physics.  Let us describe our new results starting
with the highest temperatures, where expansions in the thermal
coupling $\alpha_h=\alpha(z_h)$ are possible.  Starting from the
ultraviolet expansion of the $\beta$-function, one can solve
analytically the equations of motion in the UV.  In doing that, one
arrives at the following expansion for the metric factor up to ${\cal
O}(\alpha^2)$~\footnote{Analytical expressions up to ${\cal
O}(\alpha^3)$ will be presented in Ref.~\cite{Megias:pr}.}
\begin{equation}
b(\alpha) = \frac{\ell}{z} \left[  1 -\frac{4}{9} \beta_0 \alpha +\frac{2}{81}\left(22\beta_0^2 -9\beta_1 \right) \alpha^2  \right] \,. \label{eq:bzuv}
\end{equation}
The temperature dependence of the horizon is computed from the
second expression in Eq.~(\ref{eq:fprop}). It reads
\begin{equation}
z_h = \frac{1}{\pi T} \left[ 1 + \frac{\beta_0^2}{3}\alpha_h^2 + {\cal O}(\alpha_h^3) \right] \,. \label{eq:zh}
\end{equation}
\begin{figure}[ttt]
\begin{center}
\centerline{\includegraphics[width=7.3cm,height=5cm]{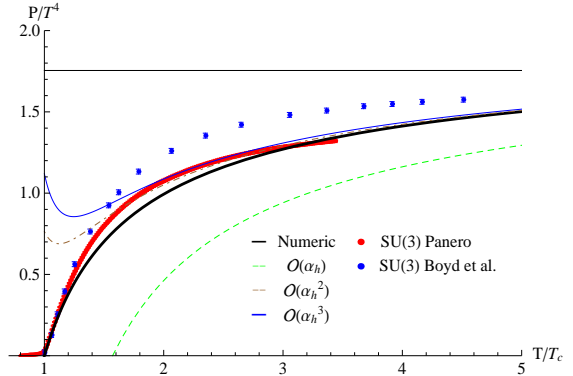}}
\end{center}
\vspace{-0.8cm}
\caption{Pressure over $T^4$ as a function of $T$ (in units of $T_c$).  We show as red points the recent lattice data for SU(3) given in Ref.~\cite{Panero:2009tv} for $N_\sigma^3 \times N_\tau = 20^3 \times 5$, and as blue points the lattice data from Ref.~\cite{Boyd:1996bx} for $N_\sigma^3 \times N_\tau = 32^3 \times 8$. The colored curves represent the analytical result from the holographic model as an expansion in powers of $\alpha_h$, c.f. Eq.~(\ref{eq:pwc}). The order ${\cal O}(\alpha_h^3)$ is computed in Ref.~\cite{Megias:pr}. The black solid line refers to the full numerical result. We have used the value of $G_5$ quoted in Eq.~(\ref{eq:G5}). }
\label{fig:pressure}
\end{figure}
\nin
\begin{figure}[ttt]
\begin{center}
\centerline{\includegraphics[width=7.3cm,height=5cm]{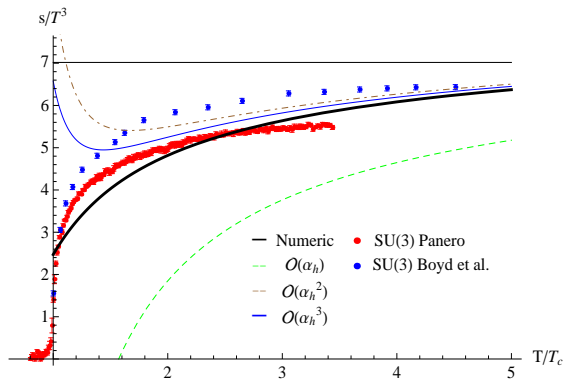}}
\end{center}
\vspace{-0.8cm}
\caption{Entropy density over $T^3$ as a function of $T$ (in units of $T_c$). See Fig.~\ref{fig:pressure} for details.}
\label{fig:entropy}
\end{figure}
\nin
\begin{figure}[ttt]
\begin{center}
\centerline{\includegraphics[width=7.3cm,height=5cm]{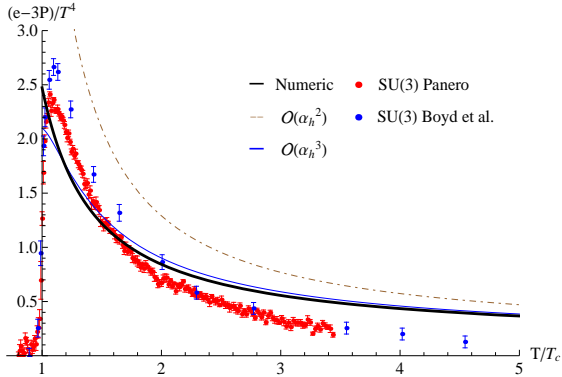}}
\end{center}
\vspace{-0.8cm}
\caption{Trace anomaly density $(\epsilon - 3p)/T^4$ as a function of $T$ (in units of $T_c$). See Fig.~\ref{fig:pressure} for details.}
\label{fig:traceanomaly}
\end{figure}

\nin The entropy density follows from the Bekenstein-Hawking entropy formula which establishes the proportionality between the entropy and the area of the event horizon of the black hole.  Up to ${\cal O}(\alpha_h^2)$ it reads
\begin{eqnarray}
&&s(T) = \frac{1}{4 G_5} b^3(z_h) \label{eq:s1text} \\
&=& \frac{\pi^3 \ell^3}{4 G_5} T^3
\Bigg[ 1 -\frac{4}{3} \beta_0 \alpha_h + \frac{1}{9} \left(
11\beta_0^2 -6\beta_1 \right) \alpha_h^2 \Bigg] \,. \nonumber
\end{eqnarray}
To reach this expression one has to evaluate Eq.~(\ref{eq:bzuv}) at the horizon, with $z_h$ given by Eq.~(\ref{eq:zh}). Using
\begin{equation}
T\frac{d\alpha_h}{dT} = -\beta_0 \alpha_h^2 - \beta_1 \alpha_h^3 + {\cal O}(\alpha_h^4) \,,
\end{equation}
it is easy to derive the pressure from Eq.~(\ref{eq:s1text}), as it is related to the entropy density by $s = dp/dT$. One gets
\begin{equation}
\frac{p(T)}{T^4} = \frac{\pi^3 \ell^3 }{16 G_5} \Bigg[ 1 - \frac{4}{3}
 \beta_0 \alpha_h + \frac{2}{9} \left( 4\beta_0^2 - 3\beta_1 \right)
 \alpha_h^2  \Bigg] \,.
\label{eq:pwc}
\end{equation}
From Eqs.~(\ref{eq:s1text}) and (\ref{eq:pwc}) one can immediately compute the energy density $\epsilon = T  s - p$, and the trace anomaly $(\epsilon - 3p)/T^4  = s/T^3 - 4p/T^4$.

In order to obtain the thermodynamic functions for all temperatures,
numerical solutions of the equations of motion have been
obtained~\cite{Megias:pr}.  We show in Figs.~\ref{fig:pressure},
\ref{fig:entropy} and \ref{fig:traceanomaly} the pressure, entropy
density and trace anamoly for pure gluodynamics with $N_c=3$. The
lattice data are taken from Refs.~\cite{Panero:2009tv,Boyd:1996bx}. We
also plot the analytical results of Eqs.~(\ref{eq:s1text}) and
(\ref{eq:pwc}). It is noteworthy and visible in the figures that the
expansion in terms of $\alpha_h$ converge quite rapidly. Even at
temperatures $T\simeq 1.5 T_c$, two orders of $\alpha_h$ give a
sufficient approximation, quite opposite to the $\alpha(T)$ expansion
in conventional high temperature QCD perturbation theory~(pQCD).

  The question arises: which value of the gravitational constant $G_5$
should we choose in the 5-dim gravity action? In principle, $G_5$ can
be chosen to reproduce the Stefan-Boltzmann limit at high
temperatures, i.e.
\begin{equation}
\frac{\pi^3\ell^3}{16 G_5^{\infty}} = (N_c^2-1)\frac{\pi^2}{45}\,. \label{eq:G5infty}
\end{equation}
If one believes that the high temperature region $T \sim (100 T_c - 1000 T_c)$ is a perturbative regime which can be described by the perturbative $\beta$-function, then this model reaches the Stefan-Boltzmann limit much slower than what lattice data suggest~\cite{Endrodi:2007tq}. This is easy to see, because the coefficient of ${\cal O}(\alpha_h)$ in the expansion of the pressure in the holographic model $(p_1^{\textrm{\tiny AdS}} = \frac{22}{3\pi})$, Eq.~(\ref{eq:pwc}), is a factor two larger than the corresponding one in pQCD $(p_1^{\textrm{\tiny pQCD}} = \frac{15}{4\pi})$~\cite{Kajantie:2002wa}. As a consequence, the value of $G_5^\infty$ given by Eq.~(\ref{eq:G5infty}) leads to lower values of the thermodynamic quantities for all temperatures in comparison with lattice data, also in the regime close to $T_c$. To reproduce lattice data in the regime $ T_c < T < 5T_c$ one must use a value of $G_5$ which is a factor $1.46$ smaller than $G_5^\infty$, i.e.
\begin{equation}
\frac{1}{G_5} = 1.46 \times \frac{1}{G_5^\infty} \,,\label{eq:G5}
\end{equation}
spoiling the Stefan-Boltzmann limit at high temperatures. Note, however, that lattice data for pressure and entropy density taken from Refs.~\cite{Panero:2009tv} and \cite{Boyd:1996bx} are not consistent each other at high temperatures, and it is hard to believe that both computations fulfill the Stefan-Boltzmann's law. This discrepancy introduces an error in $G_5$ of the order of $10\%$, which in either case is not enough to explain the factor in Eq.~(\ref{eq:G5}). The gravity model seems to be hardly consistent to reproduce at the same time lattice data at very high temperatures, and close to the phase transition. In this sense, there exists the possibility that the ideal gas limit doesn't correspond to the limit of the black hole gravity theory at high temperatures. A natural question arises: is it possible that the gravity theory allows more degrees of freedom at high temperatures? We will further address this problem, and analyze possible solutions~\cite{Megias:pr}.

We plot in the figures as a black continuus line  the numerical computation of thermodynamic quantities using for $G_5$ the value quoted in Eq.~(\ref{eq:G5}). The trace anomaly in Fig.~\ref{fig:traceanomaly} shows the characteristic decrease towards zero for high temperatures expected from pQCD. Neat $T_c$, for $p$, $\epsilon$ and $s$ in general all the AdS/QCD curves increase too slowly.

The latent heat is defined as the energy density at $T_c$, $L_h \equiv \epsilon(T_c)$. Another way to choose $G_5$ would be to reproduce the value of $L_h$ given by lattice simulations. In fact, with the value of $G_5$ quoted in Eq.~(\ref{eq:G5}) one gets  $L_h/(T_c^4 \cdot (N_c^2-1)) = 0.309$, which is in good agreement with the result from lattice, $\simeq 0.33(3)$~\cite{Panero:2009tv}. 

\begin{figure}[tbp]
\begin{center}
\centerline{\includegraphics[width=7.3cm,height=5cm]{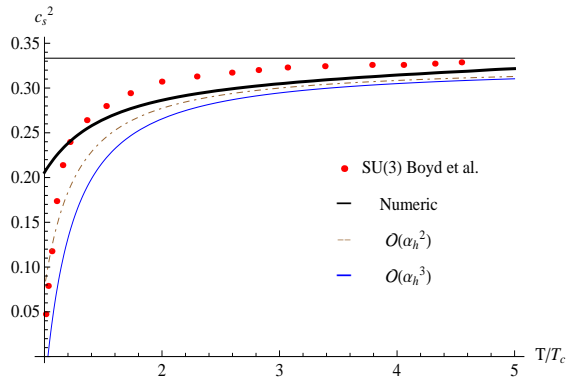}}
\end{center}
\vspace{-0.8cm}
\caption{Speed of sound squared as a function of $T$ (in units of
$T_c$). We show as points the lattice data for SU(3) taken from
Ref.~\cite{Boyd:1996bx} for $N_\sigma^3 \times N_\tau = 32^3 \times
8$. Colored curves represent the analytical result of
Eq.~(\ref{eq:cstext}) and the black solid line refers to the full
numerical computation. The order ${\cal O}(\alpha_h^3)$ is presented
in Ref.~\cite{Megias:pr}.}
\label{fig:cs2}
\end{figure}
\nin

Finally we can study the speed of sound $c_s$ which is independent of the normalization factor $G_5$. From the specific heat per unit volume $c_v = T \frac{\partial^2 p}{\partial T^2}$ and the entropy density $s$, the speed of sound writes
\begin{equation}
c_s^2 = \frac{s}{c_v} = \frac{1}{3} \Bigg[ 
 1 - \frac{4}{9}\beta_0^2 \alpha_h^2 \Bigg] \,. \label{eq:cstext}
\end{equation}
In the r.h.s. we write the result of the computation of~this quantity
in the ultraviolet from~Eqs.~(\ref{eq:s1text}) and (\ref{eq:pwc}). We
show in Fig.~\ref{fig:cs2} the speed of sound computed with the
holographic model, and compared with the lattice data of
Ref.~\cite{Boyd:1996bx}. We also show the analytical ultraviolet
approximation given by Eq.~(\ref{eq:cstext}). Since $c_s^2$ is close
to $1/3$ in the calculation, we see that we have massless excitations
in the plasma in the range $2T_c<T< 5T_c$.

\section{Conclusions}
\nin
%%%%%%%%%%%%%%%%
We study in this paper the thermodynamics of the five dimensional
dilaton-gravity model by using a parameterization for the dilaton
potential which was quite successful to reproduce the heavy $Q\bar{Q}$
potential at zero temperature~\cite{Galow:2009kw}. We~compute
analytical expressions for the pressure, entropy density and speed of
sound, as an expansion in powers of the running coupling. This
expansion turns out to converge quite rapidly even at temperatures $T
\simeq 1.5 T_c$, quite opposite to the conventional QCD perturbation
theory at high temperature.  The gravity model with the dilaton
potential of \cite{Galow:2009kw} cannot reproduce both the low
temperature $T \approx T_c$ and the Stefan-Boltzmann limit at very
high temperatures.  This finding differs from the result obtained with
a different dilaton potential shown in Ref. \cite{Gursoy:2009jd}. Both
potentials are based on $\beta$-functions which agree in leading
order. The underlying $\beta$-function of Ref.~\cite{Gursoy:2009jd}
has a coefficient $\beta_2 \alpha^4$ which is rather
large. Consequently the physics based on this model shows strong
nonperturbative features already at very small coupling. It is well
known that the coefficient $\beta_2$ is scheme dependent and it is
possible to set the physical scale for the thermodynamics also in this
model from the $Q{\bar Q}$-potential. The mapping, however, between
the $\alpha$ obtained from the gravity theory and
$\alpha_{\overline{\textrm{\tiny MS}}}$ has not been achieved. The
thermal couplings $\alpha(z_h)$ in the gravity model of
Ref.~\cite{Gursoy:2009jd} are very much smaller than the $\alpha(T)$
in the corresponding QCD-lattice simulations, thereby making the
calculation compatible with the asymptotic Stefan-Boltzmann
pressure. The swift change from a perturbative to a nonperturbative
$\beta$-function facilitates the steep rise of the thermodynamic
functions at low temperatures.

The description of Ref.~\cite{Galow:2009kw} does well for all
quantities which are calculated in the string framework, like the free
energy of a $Q \bar Q$- pair.  In a forthcoming work~\cite{Megias:pr}
we will demonstrate the agreement between the computation of the free
energy from the Bekenstein-Hawking entropy formula presented here, and
the method followed in Ref.~\cite{Gursoy:2008za} based on the
regularization of the Einstein-Hilbert action of
Eq.~(\ref{eq:action5D}). We will also extend the computation to other
thermodynamic observables, like the Polyakov loop and the spatial
string tension.
%%%%%%%%%%%%%%%%%%%%%%%%%%%
\section*{Acknowledgements}
\nin
%%%%%%%%%%%%%%%%
E.M. would like to thank the Humboldt Foundation for their stipend. This work was also supported by the ExtreMe Matter Institute EMMI in the
framework of the Helmholtz Alliance Program of the Helmholtz Association. We thank M.~Panero for providing us with the lattice data of Ref.~\cite{Panero:2009tv}.


%% References


%\bibliographystyle{elsarticle-num}


%%%%%%%%%%%%%%%%%%%%
%\vfill\eject

%\input{bib_sample}

%\bibliographystyle{h-physrev3}
%\bibliography{Refs}

\end{document}